\documentclass[prb,floatfix,twocolumn,amsmath,groupedaddress,showpacs]{revtex4}
\usepackage{bm}
\usepackage{amsmath}
\usepackage{dcolumn}
\usepackage[dvips]{graphicx}
\usepackage{bbold}
\usepackage{booktabs}
\usepackage{times}

\begin{document}
\bibliographystyle{apsrev}

%\title{
%On the Nature of the  Quantum Phase Transition in the Sandvik  Model}

\title{Quantum Phase Transition in the Four-Spin Exchange Antiferromagnet}

\author{Valeri N. Kotov}
\affiliation{Department of Physics, Boston University, 590 Commonwealth Avenue, Boston, MA 02215}
\author{Dao-Xin Yao} 
\affiliation{Department of Physics, Boston University, 590 Commonwealth Avenue, Boston, MA 02215}
\author{A. H. Castro Neto}
\affiliation{Department of Physics, Boston University, 590 Commonwealth Avenue, Boston, MA 02215}
\author{D. K. Campbell}
\affiliation{Department of Physics, Boston University, 590 Commonwealth Avenue, Boston, MA 02215}

%\date{\today}
\begin{abstract}
We study the S=1/2 Heisenberg antiferromagnet  on a square lattice
with nearest-neighbor and plaquette four-spin exchanges (introduced by A.W. Sandvik,
 Phys. Rev. Lett. {\bf 98}, 227202 (2007).)
 This model undergoes a
quantum phase transition from a spontaneously dimerized phase to N\'eel order
at a critical coupling. We show that as the critical point is approached from
 the dimerized side, the system exhibits strong fluctuations in the dimer background,
  reflected in the presence of a low-energy singlet mode, with a simultaneous rise in
 the triplet quasiparticle density. We find that  both singlet and triplet modes
 of high density condense at the transition, signaling restoration of lattice symmetry.
  In our approach, which goes beyond mean-field theory in terms of the triplet excitations,
the transition appears sharp; however since our method breaks down near the 
 critical point, we argue that we cannot make a definite conclusion regarding
 the order of the transition.
\end{abstract}


\maketitle

\section{Introduction}
Problems related to quantum criticality in quantum spin systems are of both fundamental
 and practical importance \cite{sachdevbook}. Numerous materials, such as Mott insulators, 
exhibit either  antiferromagnetic (N\'eel) order
 or quantum disordered (spin gapped) ground state    
depending on the distribution of Heisenberg exchange couplings and
 geometry. External perturbations (such as doping or frustration)
can also cause quantum transitions between these phases.
Systems with spin $1/2$ are indeed the most interesting as they are
 the most susceptible to such transitions. 
It is well  understood that  the quantum transition between
 a quantum disordered   and a N\'eel phase  is in the $O(3)$ universality
 class \cite{sachdevbook}, where a  triplet state condenses at the quantum critical point (QCP).

 A recent exciting  development in our theoretical understanding of QCPs originated
 from the proposal that if the quantum disordered (QD) phase spontaneously breaks lattice
 symmetries ({\em e.g.} is characterized by spontaneous dimer order), 
 and the transition is of second order, then exactly at the QCP spinon deconfinement occurs,
 {\em i.e.} the excitations are fractionalized \cite{senthil04s}. 
 It is assumed that the Hamiltonian itself does not break the lattice symmetries
 ({\em i.e.} does not have ``trivial'' dimer order caused by some exchanges being
 stronger than the others). We use the terms ``dimer order'' and
 ``valence bond solid (VBS) order''  interchangeably. It is expected that the  dimer order vanishes
  exactly at the point where N\'eel order appears, {\em i.e.}
 there is no coexistence between the two phases. Deconfinement thus is 
 intimately related to disappearance of VBS order; indeed if the latter
 persisted in the N\'eel phase it would be impossible to isolate
 a spinon, as ``pairing'' would always take place.
 Spontaneous VBS order driven by frustration has been a common theme in quantum
 antiferromagnetism \cite{subirread91}, although its presence  and
 the nature of criticality in specific models, such as the 2D square-lattice  frustrated Heisenberg   
 antiferromagnet, is still somewhat controversial  \cite{J1J2}.
It  would be particularly useful to apply unbiased  numerical approaches,
 such as the Quantum Monte Carlo (QMC) method,  to study  frustrated spin models; 
 however  due to the fatal ``sign" problem \cite{henelius}, frustrated Heisenberg systems
 are beyond the QMC reach.

In a recent study, the QMC method was applied to a four-spin
exchange quantum spin model without frustration, which was shown to  exhibit  
columnar dimer VBS order and
 a magnetically ordered phase with a deconfined QCP separating them \cite{sandvik0611}.
 These conclusions were later confirmed by further QMC studies \cite{melko}.
Extensions of the model, which include  for example 
additional (six-spin) interactions, provide additional
 support for a continuous QCP  \cite{lou}. 
 A different VBS pattern (plaquette order) was also proposed for the four-spin exchange model
 \cite{isaev}.
 At the same time, the nature of the quantum phase transition was   challenged
in Refs.[\onlinecite{kuklov,jiang}], where arguments were given that the transition
 is in fact of  (weakly) first order.

 It is the objective of the present work to study the Sandvik model  \cite{sandvik0611}, 
 by approaching the quantum
 transition from the dimer VBS phase. Our approach uses as a starting point a symmetry broken state
 ({\em i.e.} one out of four degenerate VBS configurations), and we thus must search
 for signatures that the system attempts to restore the lattice symmetry  at the QCP. 
 Even though full restoration is impossible within the present framework,
 we find a QCP characterized by condensation of triplet modes
 of high density;  this is in contrast to the conventional
 situation when the condensing particles are in the dilute Bose gas limit.
 The high density itself is due to the presence of a
 singlet mode that condenses at the QCP, and reflects 
the strong fluctuations of the background dimer order.
 The above effects lead to the vanishing of the VBS
 order parameter; at the same time our method, which accounts
 for the strong fluctuations, leads to a rather sharp phase
 transition. It appears that we cannot draw a definite conclusion about 
 the order of the transition because in the vicinity of the  QCP the triplon density
 increases uncontrollably, suggesting that other states (such a plaquette states and larger clusters)
 are strongly admixed into the ground state. This is generally expected in a situation
 where the lattice symmetry is restored at the quantum critical point.

%%%%%%%%%%%% FIGURE %%%%%%%%%%%%%%%%%%%%%%%%%%%%%%%%%%%
\begin{figure}[!t]
\begin{center}
\resizebox*{0.8\columnwidth}{!}{\includegraphics{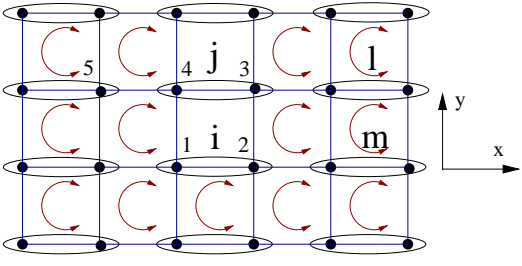}}
\end{center}
\caption{(Color online) Dimer pattern in the quantum disordered (VBS) phase, 
 $K/J > (K/J)_{c}$. }
\label{Fig1}
\end{figure}
%%%%%%%%%%%%%%%%%%%%%%%%%%%%%%%%%%%%%%%%%%%%%%%%%%%%%%%

The model under consideration is
\begin{equation}
\label{model}
H = J \sum_{\langle a,b \rangle} {\bf{S}}_{a}.{\bf{S}}_{b}
-K \sum_{a,b,c,d} ({\bf{S}}_{a}\cdot{\bf{S}}_{b})({\bf{S}}_{c}\cdot{\bf{S}}_{d}),
\end{equation}
where $J>0,K>0$, and all spins are $S=1/2$. Consider the numbers 1,2,3,4 
in Fig.~\ref{Fig1}. The summation in the four-spin term is
 over indexes $(a,b)=(1,2),(c,d)=(3,4)$ and  $(a,b)=(1,4),(c,d)=(2,3)$
 on a given plaquette, and then
 summation is made over all plaquettes \cite{lauchli05}. The range of parameters
 explored in Ref.[\onlinecite{sandvik0611}] is $K/J \leq 2$,  and the  QCP  is at 
 $(K/J)_{c} \approx 1.85$.
Our coupling notation is slightly different from  the one used in Refs.[\onlinecite{sandvik0611,melko}]; 
 the  coupling $K$ is related to the parameter  $Q$ \cite{sandvik0611,melko} via $K = Q/(1+ Q/(2J))$,
 and the critical point in that notation is  $(Q/J)_{c} \approx 25$.
The dimerization pattern is proposed to
 be of the ``columnar'' type, as shown in Fig.~\ref{Fig1}. Four such configurations
  exist.  We will  assume a configuration of this type, 
 will  show that it is stable  at $K/J \gg 1$, and will then search
 for an instability towards the N\'eel state as $K/J$ decreases.

 The rest of the paper is organized as followed.  In Section II we present results based on
the mean-field approach in terms of
 the dimer (triplon) operators. In Section III we extend our treatment beyond mean-field,
 and even further in Section IV, where we also take into account low-energy singlet two-triplon excitations.
 Section V contains our conclusions.

\section{Mean-field treatment}
%{\it Mean-field treatment.} $-$ 
We start by rewriting  Eq.~(\ref{model})  in  the
the bond-operator representation \cite{sachdev90}, where on a dimer ${\bf i}$, the two spins
forming it are expressed as:
$S_{1,2}^{\alpha} = \frac{1}{2} ( \pm s_{\bf i}^{\dagger}  t_{{\bf i}{\alpha}} \pm
t_{{\bf i}{\alpha}}^{\dagger} s_{\bf i}  -
i \epsilon_{\alpha\beta\gamma} t_{{\bf i}{\beta}}^{\dagger}
t_{{\bf i}{\gamma}})$,  
%\begin{equation}
%\label{bondops}
%S_{1,2}^{\alpha} = \frac{1}{2} ( \pm s_{\bf i}^{\dagger}  t_{{\bf i}{\alpha}} \pm
%t_{{\bf i}{\alpha}}^{\dagger} s_{\bf i}  - 
%i \epsilon_{\alpha\beta\gamma} t_{{\bf i}{\beta}}^{\dagger}
%t_{{\bf i}{\gamma}}),
%\end{equation}
and $s_{\bf i}^{\dagger}, t_{{\bf i}{\alpha}}^{\dagger}, \alpha=x,y,z$
create a singlet and triplet of states.
 We refer to the triplet (S=1) quasiparticle, $t_{{\bf i}{\alpha}}^{\dagger}$,  as ``triplon".
The bold indexes ${\bf i},{\bf j},{\bf m},{\bf l}$ label the  dimers (see Fig.~\ref{Fig1}).
Summation over repeated Greek indexes is  assumed, unless indicated otherwise.
%\noindent

The hard-core constraint, $s_{\bf i}^{\dagger}s_{\bf i}+t_{{\bf i}{\alpha}}^{\dagger}t_{{\bf i}{\alpha}}=1$, 
 must be enforced on every site, which at the mean-field (MF)
 level can be done by introducing a term in the Hamiltonian,
$-\mu \sum_{{\bf i}} (s^2 + t_{{\bf i}{\alpha}}^{\dagger}t_{{\bf i}{\alpha}}-1)$. Then  
  $\mu$ and the (condensed) singlet amplitude $s\equiv \langle s_{\bf i}\rangle$,
 are determined by the MF equations \cite{sachdev90}.  
 We obtain at the quadratic level, in momentum representation:
\begin{equation}
\label{quadratic}
H_{2} = \sum_{\bf{k}, \alpha} \left \{ A_{\bf{k}} t_{\bf{k}\alpha}^{\dagger}t_{\bf{k}\alpha}
+ \frac{B_{\bf{k}}}{2}\left(t_{\bf{k}\alpha}^{\dagger}
 t_{\bf{-k}\alpha}^{\dagger}
+ \mbox{h.c.}\right) \right \}
\end{equation}
where
\begin{eqnarray}
\label{AB}
A_{\bf{k}}& =& J/4 - \mu + s^{2}( \xi_{\bf k}^{-} + K/2) + s^{4} \Sigma({\bf k}) \ ,
\nonumber \\
B_{\bf{k}}& = &s^{2} \xi_{\bf k}^{+}  +  s^{4} \Sigma({\bf k}) \ ,  \nonumber \\
\xi_{\bf k}^{\pm}& =&  -(J/2) \cos{k_{x}} +  ( J\pm K/4 ) \cos{k_{y}}  \ .  
\end{eqnarray}

\begin{figure}[!t]
\begin{center}
\resizebox*{0.95\columnwidth}{!}{\includegraphics{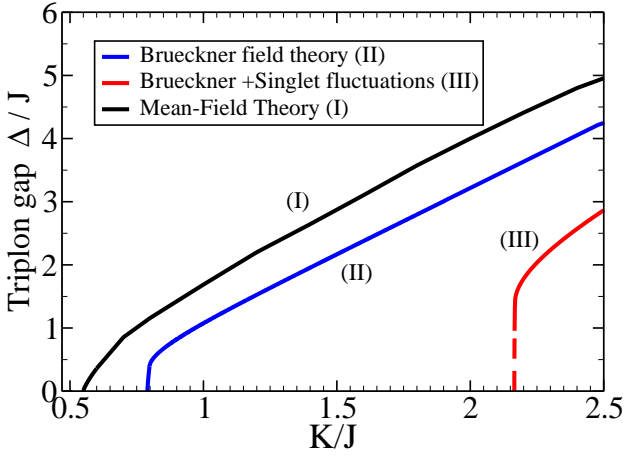}}
\end{center}
\caption{(Color online) Triplon excitation gap $\Delta = \omega({\bf k}_{AF})$ in various approximations.
The point $\Delta \rightarrow 0$ corresponds to  transition to the  N\'eel phase.}
\label{Fig2}
\end{figure}

The four-spin interaction from (\ref{model}) acting  between two dimers
({\em e.g.} ${\bf i},{\bf j}$ in Fig.~\ref{Fig1}) contributes to the ``on-site'' gap and hopping
 ($\xi_{\bf k}^{-}$) via  $A_{\bf{k}}$, as well as to 
the  quantum fluctuations term $B_{\bf{k}}$.  The part involving
 four dimers has been split in a mean-field fashion, leading to the
Hartree-Fock self-energy
\begin{equation}
\label{selfenergyK}
-\Sigma({\bf k})/K = 2 \Sigma_x \cos{k_{x}} +2\Sigma_y \cos{k_{y}}
+ \Sigma_{xy} \cos{k_{x}}\cos{k_{y}},
\end{equation}
with 
%$\Sigma_x = (1/3) \sum_{\alpha} \langle t_{{\bf i}\alpha}^{\dagger}t_{{\bf m}\alpha}
%+t_{{\bf i}\alpha}^{\dagger}t_{{\bf m}\alpha}^{\dagger} \rangle$,
\begin{equation}
\Sigma_x = \frac{1}{3} \sum_{\alpha} \langle t_{{\bf i}\alpha}^{\dagger}t_{{\bf m}\alpha}
+t_{{\bf i}\alpha}^{\dagger}t_{{\bf m}\alpha}^{\dagger} \rangle \ ,
\end{equation}
where ${\bf i},{\bf m}$ are neighboring dimers in the $x$ (horizontal) direction (Fig.~\ref{Fig1}),
 and similarly for the $y$ and the diagonal contributions.
The triplon dispersion is $\omega({\bf k})= \sqrt{A_{\bf{k}}^{2} - B_{\bf{k}}^{2}}$,
 and has a minimum at the N\'eel ordering wave-vector ${\bf k}_{AF}=(0,\pi)$
(since we work on a dimerized lattice).
The ground state energy is then easily computed, 
\begin{equation}
E_{GS} = E_{0} + \langle H_{2} \rangle \ ,
\end{equation}
where
\begin{eqnarray}
 E_{0}/N &=& -\frac{3}{4} (Js^{2} + Ks^{4}) +\mu (-s^{2}+1) +\\ 
&&3Ks^{4}\left (\Sigma_x^{2}
+ \Sigma_y ^{2} + \frac{1}{2}  \Sigma_{xy}^{2} \right ) ,
\end{eqnarray}
and
\begin{equation}
 \langle H_{2} \rangle = \frac{3}{2}\sum_{\bf{k}}\left ( \omega({\bf k}) -A_{\bf{k}} \right ) .
\end{equation}
 The mean-field equations
 require a numerical  minimization of $E_{GS}$ with respect to the parameters $\{\mu,s,\Sigma_x,\Sigma_y,
\Sigma_{xy} \}$. This amounts to  the self-consistent Hartree-Fock  approximation 
 for $\Sigma({\bf k})$. The result for the triplon gap $\Delta = \omega({\bf k}_{AF})$
is presented in  Fig.~\ref{Fig2} (black curve).

The MF result  $(K/J)_{c} \approx 0.6$ substantially underestimates the location of the
critical point, compared to the  the QMC calculations, where $(K/J)_{c} \approx 1.85$ \cite{sandvik0611,melko}.
Interestingly, if one solves the MF equations ignoring
both the hard core and the $\Sigma({\bf k})$, one finds $(K/J)_{c}=1$. Physically, in
the  full MF, the hard core contribution increases the gap (and hence
the stability of the dimer phase) while at the same time suppressing the
antiferromagnetic fluctuations (which favor the N\'eel state).

We also note that a recent (hierarchical) MF treatment based on the plaquette ground state also
underestimates very strongly the QCP location ($(K/J)_{c} \approx 1$ \cite{isaev}), similarly to our result.
In our view this means that both mean field approaches are not sufficient to attack the present
 problem, where fluctuations are  apparently  very strong. We choose to accept that the numerical QMC
 result gives the most accurate determination of the QCP, and therefore in what follows we extend 
our treatment in several directions beyond mean-field theory.

%%%%%%%%%%%% FIGURE %%%%%%%%%%%%%%%%%%%%%%%%%%%%%%%%%%%
%%%%%%%%%%%%%%%%%%%%%%%%%%%%%%%%%%%%%%%%%%%%%%%%%%%%%%%

\section{Beyond mean-field: The Dilute triplon gas approximation}
%{\it Beyond mean-field. Dilute triplon gas approximation.} $-$
A more accurate treatment of fluctuations is possible by taking into account the hard-core
 constraint beyond mean-field. One can set  the singlet amplitude $s=1$ in the previous
 formulas, but 
  introduce an infinite on-site repulsion between the triplons, 
$U \sum_{{\bf i},\alpha \beta} t_{\alpha {\bf i} }^{\dagger}t_{\beta {\bf i}}^{\dagger}t_{\beta {\bf i}}
t_{\alpha {\bf i}}, \ U \rightarrow \infty$. As long as the triplon density (determined by the
 quantum fluctuations) is low, an infinite repulsion corresponds
 to a finite scattering amplitude between excitations and can be 
calculated by resumming ladder diagrams for the  scattering vertex \cite{fetterbook}.
 This leads to the effective triplon-triplon vertex $\Gamma({\bf{k}}, \omega)$ which was
 previously calculated \cite{kotov98}:
\begin{equation}
\label{HCvertex}
\Gamma^{-1}({\bf{k}}, \omega) \!   =  \!
\sum_{\bf{q}} \frac{u_{\bf{q}}^{2}
u_{\bf{k}- \bf{q}}^{2}}{\omega(\bf{q}) + \omega(\bf{k}- \bf{q}) -\omega}
 + \left\{ \begin{array}{c} u \rightarrow v \\
\omega \rightarrow -\omega \end{array} \right\}.
\end{equation}
This vertex in turn affects the triplon dispersion via 
(what we call) the Brueckner self-energy \cite{kotov98}:
\begin{equation}
\label{Br}
\Sigma_{B}({\bf{k}},\omega) = 4\sum_{\bf{q}} v_{\bf{q}}^{2}
\Gamma({\bf{k}} + \bf{q}, \omega - \omega(\bf{q})).
\end{equation}
The corresponding parameters in the quadratic Hamiltonian (\ref{quadratic}) in this case are
\begin{eqnarray}
\label{AB1}
A_{\bf{k}}& =& J + 2K(1-4n_{t}/3)+
\xi_{\bf k}^{-} + \Sigma({\bf k}) + \Sigma_{B}({\bf{k}},0),
\nonumber \\
B_{\bf{k}}& = & \xi_{\bf k}^{+}  + \Sigma({\bf k}).
\end{eqnarray}
The Bogolubov coefficients are defined in the usual way
 $u_{\bf{k}}^{2} = 1/2 + A_{\bf{k}}/(2\omega({\bf k}))=1+v_{\bf{k}}^{2}$. 
The various terms in $\Sigma({\bf k})$ can be expressed through them:
 for example $\Sigma_x = \sum_{\bf k}(v_{\bf{k}}^{2} + v_{\bf{k}}u_{\bf{k}})\cos{k_{x}}$, and
 so on. The density of triplons is $n_t = \langle t_{{\bf i}\alpha}^{\dagger} t_{{\bf i}\alpha}\rangle
= 3 \sum_{\bf k}v_{\bf{k}}^{2}$.
In addition, the renormalization of the quasiparticle residue, 
 $Z^{-1}_{\bf k} = 1- \partial \Sigma_{B}({\bf{k}},0)/\partial \omega$, 
 implies the replacement $u_{\bf{k}} \rightarrow \sqrt{Z_{\bf k}}u_{\bf{k}}, \ 
v_{\bf{k}} \rightarrow \sqrt{Z_{\bf k}}v_{\bf{k}}$ in all the formulas \cite{kotov98}, and the 
renormalized spectrum $\omega({\bf k}) = Z_{\bf k} \sqrt{A_{\bf{k}}^{2} - B_{\bf{k}}^{2}}$. 

  %%%%%%%%%%%%%%%%%%%%%%%%%%%%%%%%%%%%%%%%%%%%%%%%%%%%%%%
\begin{figure}[!t]
\begin{center}
\resizebox*{0.55\columnwidth}{!}{\includegraphics{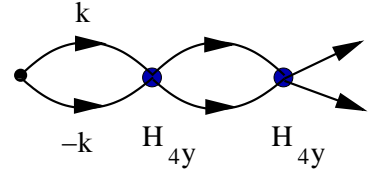}}
\end{center}
\caption{Renormalization of quantum fluctuations by  resummation of 
 a ladder series, with (\ref{interactionsy}) at the vertices.}
\label{Fig3}
\end{figure}
%%%%%%%%%%%%%%%%%%%%%%%%%%%%%%%%%%%%%%%%%%%%%%%%%%%%%

 An iterative numerical evaluation of the spectrum using the above equations,
 which amounts to solution of the Dyson equation, leads to the result shown in Fig.~\ref{Fig2} 
(blue curve).
 The above approach appears to be well justified since the quasiparticle
 density $n_t < 0.1$.  The resulting critical  point is still in the ``weak-coupling'' regime $K/J < 1$,
 with about 100\% deviation from the QMC result ($(K/J)_c \approx 1.85$). This suggests
 that the on-site triplon fluctuations  are not the dominant cause for the disagreement
 with the QMC results; thus we proceed to include two-particle fluctuations
 (in the triplon language), which amounts to including dimer-dimer correlations. 

\section{ Strong fluctuations in the singlet background: QCP beyond the dilute triplon gas approximation}
%{\it Strong fluctuations in the singlet background.
 %QCP beyond the dilute triplon gas approximation.} $-$
It is clear that ``non-perturbative'' effects are responsible for driving the
QCP towards the ``strong-coupling'' region $ K/J \sim 2$. To proceed we make two
improvements to the previous low-density, weak-coupling theory.

 First, we take into account fluctuations in the singlet background,
 {\em i.e.} the manifold on which the triplons are built and interact.
  The main effect originates from the action of the four-spin $K$-term from
 (\ref{model}) on two dimers, {\em e.g.} ${\bf i},{\bf j}$ in Fig.~\ref{Fig1}.
Part of this action has led to the on-site gap $2K$ in (\ref{AB1}), favoring
 dimerization. However, a strong attraction between the two dimers is also
 present,  since the $K$-term is symmetric with respect to the index pair exchange
 $(1,2)(3,4) \leftrightarrow(1,4)(2,3)$, leading to a ``plaquettization''  tendency
as well. In the triplon language this is manifested by formation of bound
 states of two triplons, due to their nearest-neighbor interactions
\begin{eqnarray}
\label{interactionsy}
H_{4,y}&=& \sum_{\langle {\bf{i,j}} \rangle_{y},\alpha\beta}
\left\{\gamma_{1} t_{\alpha {\bf i}}^{\dagger}t_{\beta {\bf j}}^{\dagger}t_{\beta {\bf i}}
t_{\alpha {\bf j}} +\gamma_{2} t_{\alpha {\bf i}}^{\dagger}t_{\alpha {\bf j}}^{\dagger}
t_{\beta {\bf i}}t_{\beta {\bf j}}   \right. \nonumber  \\
&& 
\left. + \ \gamma_3  t_{\alpha {\bf i}}^{\dagger}t_{\beta {\bf j}}^{\dagger}
t_{\alpha {\bf i}}t_{\beta {\bf j}} \right\},  \\
% \ \gamma_{1}&=&-\frac{K}{8} + \frac{J}{2}, \ \gamma_{2}=-\frac{K}{8} - \frac{J}{2}, \
%\gamma_3 = -\frac{5K}{4}. \nonumber\\
\gamma_{1}&=&-\frac{K}{8} + \frac{J}{2}, \  \gamma_{2}=-\frac{K}{8} - \frac{J}{2}, \ 
\gamma_3 = -\frac{5K}{4}. \nonumber 
\end{eqnarray}
We  also checked that
 on the perturbative (Hartree-Fock) level, the effect of this
 term on equations (\ref{AB}) and (\ref{AB1}) was  negligible
(and we did not write it explicitly).

%%%%%%%%%%%% FIGURE %%%%%%%%%%%%%%%%%%%%%%%%%%%%%%%%%%%
\begin{figure}[!t]
\begin{center}
\resizebox*{0.96\columnwidth}{!}{\includegraphics{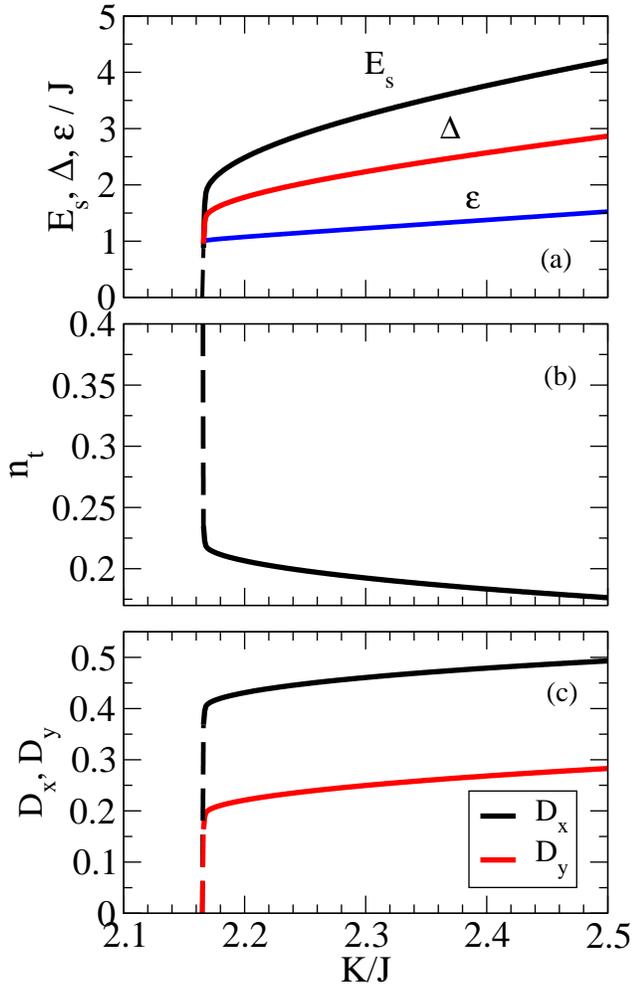}}
\end{center}
\caption{(Color online) (a.) Singlet bound state energy $E_s$ (black), 
 binding energy $\epsilon = 2\Delta - E_s$ (blue), and  the  triplon gap $\Delta$
(red). (b.) Triplon density $n_t$. (c.) Dimer order parameters.
Dashed parts of the lines represent points corresponding to rapid growth
 of the quasiparticle density.}
\label{Fig4}
\end{figure}
%%%%%%%%%%%%%%%%%%%%%%%%%%%%%%%%%%%%%%%%%%%%%%%%%%%%%%%

An intuitive way of taking into account the effect of two-triplon
 bound states (with total spin S=0) on the one-triplon spectrum,
 is to work in the ``local'' approximation. This means  effectively
 neglecting the triplon dispersion  and directly evaluating  
the  ladder series that renormalizes the quantum fluctuation
 term $B_{\bf{k}}$ in (\ref{quadratic}), corresponding to
 emission of a pair of triplons with zero total momentum.
This is illustrated graphically in Fig.~\ref{Fig3}, with the result
\begin{eqnarray}
B_{\bf{k}} &=&  -\frac{J}{2} \cos{k_{x}} +  \left (\frac{J+ K/4}{1- \frac{|\gamma|}{|\Delta E|}}
\right ) \ \cos{k_{y}} 
+\Sigma({\bf k}) \nonumber \ , \\
&& \gamma \equiv \gamma_1 + 3 \gamma_2 + \gamma_3 = -J - \frac{7}{4}K,
\end{eqnarray}
where $\gamma$ is the effective attraction of two triplons  with total $S=0$, and
\begin{equation} 
\Delta E = 2J + \frac{11}{4}K 
\end{equation}
 is the energy of two (non-interacting) triplons
on adjacent sites. This calculation is  justified
 for $K/J \gg 1$ and leads to an increase of the quantum fluctuations, 
 and from there to almost doubling of the  triplon density $n_t$ (see Fig.~\ref{Fig4} below).
It contributes significantly to the shift of the QCP.

We can go beyond the ``local" approximation by solving the Bethe-Salpeter equation for the bound
 state, formed due to the attraction (\ref{interactionsy}), 
 and taking into account the full triplon dispersion.
 The equation for the singlet bound state energy $E_{s}({\bf Q})$, 
 corresponding to total pair momentum ${\bf Q}$ is
\begin{equation}
1   =  2 \gamma \sum_{{\bf q}} \!
 \frac{u_{{\bf q}}^{4} \cos^{2}{q_{y}}}{E_{s}({\bf Q})
-\omega({\bf Q}/2 + {\bf q})-\omega({\bf Q}/2 - {\bf q})}.
\end{equation}
Here we have, for simplicity, written only the main contribution to pairing (Eq.~(\ref{interactionsy})) 
 in the limit $K/J \gg 1$, and have neglected the on-site repulsion (which leads
 to slightly diminished pairing), as well as small pairing due to
 the exchange $J$ from dimers in the $x$-direction on Fig.~\ref{Fig1}.
 It is easily seen that the lowest energy corresponds to ${\bf Q}=0$;
we define from now on $E_{s} \equiv E_{s}({\bf Q}=0)$.
The binding energy is $\epsilon = 2\Delta - E_{s}$,
  where $\Delta$ is the one-particle gap.
 The bound state wave-function corresponding to $E_{s}$ is
$|\Psi \rangle = \sum_{\alpha,{\bf i}, {\bf j},{q_y}} \Psi_{q_y} e^{i q_y ({\bf i} -  {\bf j})}  
t_{\alpha {\bf i}}^{\dagger} t_{\alpha {\bf j}}^{\dagger}|0\rangle$.
 In the ``local" limit (nearest-neighbor pairing), $\Psi_{q_y}=\sqrt{2} \cos{q_y}$.

Second, we have  made subtle changes to the resummation
 procedure concerning the quasiparticle renormalization $Z$,
 based on both formal and physical grounds. 
 On the one hand it is clear that in the Brueckner approximation
 (Eq.~(\ref{Br})), where the self energy is linear in the 
density ($\Sigma_{B} \propto n_t$), the dependence of the vertex $\Gamma^{-1}$
on density is beyond the accuracy of the calculation, meaning
 one can put $u_{\bf q}=1, v_{\bf q}=0$  in (\ref{HCvertex}), instead of
 determining them self-consistently. This leads to a decreased
influence of the hard-core  $\Sigma_{B}$ (which favors the dimer state)  
 on the Hartree-Fock self-energy $\Sigma({\bf k})$ from (\ref{selfenergyK})
(which favors the N\'eel state). It is  indeed the mutual interplay
 between $\Sigma_{B}>0$ and $\Sigma({\bf k})<0$, that determines the exact location
 of the QCP in the course of the Dyson's equation iterative solution.
 While in the ``weak-coupling'' regime $K/J<1$, $\Sigma_{B}$ always dominates,
 in the ``strong-coupling'' region $K/J > 2$, $\Sigma({\bf k})$ starts playing
 a significant role, since parametrically $ \Sigma \propto K n_t$. 
 It is physically consistent that in the region where singlet fluctuations
 in the dimer background are strong, the hard-core effect is less important, 
{\em i.e.} in effect the kinematic hard-core constraint is ``relaxed''. 
% we have in fact chosen to neglect completely the $Z$ renormalization
%$ of $u_{\bf k},v_{\bf k}$, although we have found that the results are quite generic).
 We also observe that in typical models with QCP driven by explicit dimerization, 
such as the bilayer model, the described   difference in 
 approximation schemes makes a very small difference on the location of
 the QCP \cite{shevchenko00}, since those models are always in the
 ``weak-coupling'' regime, dominated by the hard-core repulsion
 of excitations on a  non-fluctuating dimer configuration.
 The purpose of the above rather technical diversion is to emphasize 
that  care has been taken to take into account as accurately as possible
 the effect of the (low-energy) two-particle spectrum on the one-particle triplon gap.

 Our results are summarized in Fig.~\ref{Fig4} and  Fig.~\ref{Fig2} (red line)
for the gap. The critical point is shifted towards $(K/J)_{c} \approx 2.16$
(in much better agreement with QMC data), with a very strong increase of the density towards $K_c$.
This translates into a decrease of the dimer order, as measured by the two dimer
 order parameters that we compute from the expressions:
$D_{x} = |\langle {\bf S}_3 \cdot {\bf S}_4 \rangle - \langle {\bf S}_5 \cdot {\bf S}_4 \rangle |
= |-\frac{3}{4} + n_t + \frac{3}{4}\Sigma_x|$,
 $D_{y} = |\langle {\bf S}_3 \cdot {\bf S}_4 \rangle - \langle {\bf S}_1 \cdot {\bf S}_4 \rangle|
= |-\frac{3}{4} + n_t -\frac{3}{4}\Sigma_y|$. The spins are labeled as in Fig.~\ref{Fig1}. 
 The singlet bound state energy $E_s(0)$ also tends towards zero at the QCP, 
 with the corresponding binding energy  remaining quite large $\epsilon/J \approx 1$.
 All these effects point towards a tendency of the system to restore
 the lattice symmetry, although it is certainly clear that as the critical
 point is approached,  our approximation scheme (low density of quasiparticles) 
  breaks down (dashed lines on figures).  
We should point out that the sharpness of variation near $K_c$ is not due to
divergence in any of the self-energies but is a result of rapid cancellation
 at high orders ({\em i.e.} iterations in the Dyson equation).
 In fact cutting off our iterative procedure at finite order gives a smooth
 curve,  suggesting that additional classes of diagrams become
 important (although in practice their classification is an insurmountable task).
The merger of singlet and triplet modes, which we find near the QCP, 
  in principle reflects a tendency towards quasiparticle fractionalization
(spinon deconfinement) and is also found in the 1D Heisenberg chain
 with frustration \cite{Weihong}, where spinons are always deconfined.

Since we are now dealing with a situation where the density is not very small $n_t \approx 0.2$, 
 it is prudent to check how the next order in the density  may affect the above results.
For example at  second order in the density, the self-energy $\Sigma({\bf k})$ changes by
 amount $\delta \Sigma({\bf k})$, i.e. one has to add this contribution
 to the right hand side of Eq.(\ref{AB1}), namely:
\begin{equation}
A_{\bf{k}} \rightarrow A_{\bf{k}} + \delta \Sigma({\bf k}), \ \
B_{\bf{k}} \rightarrow B_{\bf{k}} + \delta \Sigma({\bf k}).
\end{equation}
 We have found
\begin{eqnarray}
\delta \Sigma({\bf k})& =& -2K (Q_{x}^{2} -  P_{x}^{2})\cos{k_{x}} + \nonumber \\
&&
+ 2K (Q_{y}^{2} -  P_{y}^{2})\cos{k_{y}} - \nonumber \\
&& -K (Q_{xy}^{2} -  P_{xy}^{2})\cos{k_{x}}\cos{k_{y}},
\end{eqnarray}
and the following definitions are used:
\begin{eqnarray}
&&P_{x} = (1/3) \sum_{\alpha} \langle t_{{\bf i}\alpha}^{\dagger}t_{{\bf m}\alpha}\rangle
= \sum_{\bf k}v_{\bf{k}}^{2}\cos{k_{x}} , \nonumber \\
&& Q_{x} = (1/3) \sum_{\alpha} \langle t_{{\bf i}\alpha}t_{{\bf m}\alpha}\rangle
= \sum_{\bf k}u_{\bf{k}}v_{\bf{k}}\cos{k_{x}}  , 
\end{eqnarray}
and similarly for the other directions, for example 
$P_{y} = \sum_{\bf k}v_{\bf{k}}^{2}\cos{k_{y}}$,
$ P_{xy} = \sum_{\bf k} v_{\bf{k}}^{2} \cos{k_{x}}\cos{k_{y}}$, etc.
After including these expressions in our numerical iterative procedure,
 we have found that the QCP is shifted by a very small amount,
 and the overall picture, as summarized in Fig.~\ref{Fig4} and  Fig.~\ref{Fig2} (red line),
 still stands.

\section{Conclusions}
In conclusion,  we have shown that the QCP between the N\'eel and the dimer
 state in the model (\ref{model}) is of unconventional nature, in the sense that it is characterized
 by the presence of both triplet and singlet low-energy modes.
Near the QCP, whose location ($(K/J)_{c}\approx 2.16$) we find in fairly good agreement
 with recent QMC studies,
 the system exhibits: (1.) Strong rise of the triplon excitation density, due to increased
 quantum fluctuations, (2.) Corresponding strong decrease (and ultimately vanishing) of the dimer order
 at the QCP  (3.) Vanishing of a singlet energy scale, related to the destruction of the dimer 
``columns" in Fig.~\ref{Fig1}.
 The above  effects are all related and influence strongly one another, ultimately meaning that 
the QCP reflects strong fluctuations and can not be described in a mean-field theory framework. 
 These results also suggest a desire of the system to restore the lattice symmetry at the QCP,
 as found in the QMC studies \cite{sandvik0611}.

 At the same time all our improvements  beyond  mean-field theory have also resulted in
 a very sharp transition, which appears to be first order. However in our view our approach is not
 capable of addressing correctly the issue of the order of the phase transition, basically because
once we take the strong (inter) dimer fluctuations in to account, the triplon density starts
rising quickly beyond control. This is in a certain sense natural in a situation where
 the system wants to restore the lattice symmetry at the QCP and thus the ground
 state acquires strong admixture of plaquette, etc. fluctuations as the dimers  begin to ``disappear."
 This is also manifested in the fact that our procedure is sensitive 
 to the number of iterations in the Dyson equation; all presented
 results are for  an ``infinite" number of iterations, so that a fixed point is reached, but
 cutting off the procedure results in a smoother behavior and a shift of the QCP, which
 becomes iteration dependent.  We have not previously  encountered such volatile behavior in any other spin model  with a dimer to magnetic order transition.
 Since iterations translate into accounting of more and more fluctuations, the sensitivity
 of the results seems to mean that the situation starts spiraling out of control  near the QCP,
 quite likely because classes of fluctuations become important that are not included in
 the dimer description, such as longer range correlations, etc. All this suggests that
 the triplon quasiparticle description breaks down  near the QCP which indeed appears natural
 in a model where spinon deconfinement is expected to take place at the QCP \cite{sandvik0611}.
On the other hand, if we put aside the arguments  that our approach is not reliable 
 near the QCP,  the natural conclusion would be that the transition is first order.

\acknowledgments
We are grateful to A. W. Sandvik, K. S. D. Beach, S. Sachdev, and  O. P. Sushkov 
for  numerous stimulating discussions.  A.H.C.N. was
supported through NSF grant DMR-0343790; V.N.K., D.X.Y., and D.K.C. were supported by
 Boston University.

%\bibliography{jk.bib}
%\bibliographystyle{forprl}

\end{document}